\documentclass[preprint,
onecolumn,
superscriptaddress,
titlepage,
nofootinbib,
amsmath,
amssymb,
aps]{revtex4-2}

\usepackage{url,hyperref,lineno,microtype}
\usepackage{journals}

\usepackage{graphicx}
\usepackage{booktabs}
\usepackage{dcolumn}
\usepackage{xcolor}

\newcommand{\ie}{i.e.,~}
\newcommand{\eg}{e.g.,~}

\newcommand{\unit}[1]{%
    \ensuremath{\, \mathrm{#1}}}

\newcolumntype{d}[1]{D{.}{.}{#1}}

\newcommand{\rrsst}{r^2\sin^2\theta}

\usepackage[T2A,T1]{fontenc}

\begin{document}

\title{Fast Rotating Neutron Stars: Oscillations and Instabilities} 

\author{Christian J. Kr\"uger}
    \email{christian.krueger@tat.uni-tuebingen.de}
    \affiliation{Theoretical Astrophysics, IAAT, University of T\"ubingen, 72076 T\"ubingen, Germany}
\author{Kostas D. Kokkotas}
    \email{kostas.kokkotas@uni-tuebingen.de}
    \affiliation{Theoretical Astrophysics, IAAT, University of T\"ubingen, 72076 T\"ubingen, Germany}
\author{Praveen Manoharan}
    \email{praveen.manoharan@uni-tuebingen.de}
    \affiliation{Theoretical Astrophysics, IAAT, University of T\"ubingen, 72076 T\"ubingen, Germany}
\author{Sebastian H. V\"olkel}
    \email{sebastian.voelkel@sissa.it}
    \affiliation{SISSA, Via Bonomea 265, 34136 Trieste, Italy and INFN Sezione di Trieste}
    \affiliation{IFPU - Institute for Fundamental Physics of the Universe, Via Beirut 2, 34014 Trieste, Italy}

\date{\today}

\begin{abstract}
In this review article, we present the main results from our most recent research concerning the oscillations of fast rotating neutron stars. We derive a set of time evolution equations for the investigation of non-axisymmetric oscillations of rapidly rotating compact objects in full general relativity, taking into account the contribution of a dynamic spacetime. Using our code, which features high accuracy at comparably low computational expense, we are able to extract the frequencies of non-axisymmetric modes of compact objects with rotation rates up to the Kepler limit. We propose various universal relations combining bulk properties of isolated neutron stars as well as of binary systems before and after merger; these relations are independent of the true equation of state and may serve as a valuable tool for gravitational wave asteroseismology. We also present an introductory example using a Bayesian analysis.
\end{abstract}

\maketitle

\section{Introduction}

Oscillations and instabilities of neutron stars were always considered among the promising sources for gravitational waves. The systematic study of non-axisymmetric neutron star oscillations began in the 1960s with the pioneering works of Thorne and collaborators (cf. \cite{1967ApJ...149..591T, 1969ApJ...158....1T} (and subsequent papers)), in which they laid out the equations governing the perturbations of compact stars in general relativity. The numerical solution of these equations has proven highly challenging and it has taken nearly two decades before Lindblom and Detweiler found an advantageous formulation of the eigenvalue problem that allowed them to determine the complex frequencies of the acoustic modes of sufficiently realistic stellar models (\cite{1983ApJS...53...73L, 1985ApJ...292...12D}). These results did not conclude the investigation of these modes; in particular  \cite{1991RSPSA.432..247C, 1991RSPSA.434..449C} turned to the perturbations of relativistic stars and studied their oscillations as a scattering problem. 

In the mid-1980s, inspired by a toy model \cite{1986GReGr..18..913K}
suggested that the dynamic spacetime of a neutron star exhibits its very own class of modes and christened them $w$-modes (\cite{1992MNRAS.255..119K}). The short damping times of these modes (comparable to the ones of black-holes) pose numerical challenges, 
 but the application of the continued fraction method (\cite{1985RSPSA.402..285L}) and numerical integration along anti-Stokes lines (\cite{1995MNRAS.274.1039A}) improved considerably the accuracy in the calculation of both \emph{fluid} and \emph{spacetime} modes  and led to the discovery of the new class of $w$-modes, the so-called $w_{\rm II}$-modes (\cite{1993PhRvD..48.3467L}). This advancement in computational accuracy allowed the study of many neutron star equations of state leading to the discovery of universal relations involving frequencies and damping times of oscillation modes. These universal relations provided ways for solving the inverse problem within the so-called \emph{gravitational wave asteroseismology}, allowing to set constrains on the mass and/or radius of the neutron star, and eventually the nuclear equation of state, via the observation of oscillations (see, e.g., \cite{1996PhRvL..77.4134A, 1998MNRAS.299.1059A, 2001MNRAS.320..307K, 2004PhRvD..70l4015B}).

With increasing computational power during the 1990s, the first attempts were made to increase the dimensionality of the hitherto (due to the restriction to spherical symmetry) purely radial problem and include time as a second dimension. The first successful time evolutions of perturbations of relativistic neutron stars were reported by  \cite{1998PhRvD..58l4012A}. 
\cite{2000PhDT.......170R, 2001PhRvD..63f4018R} reformulated those evolution equations by means of the ADM-formalism (\cite{ADMformalism}) and  introduced a non-uniform radial grid to ensure stable numerical evolution  when using realistic equations of state .

%%%%%%%%
\subsection{Asteroseismology}
%%%%%%%%

The different oscillation patterns of a neutron star are characterized by their restoring force, \eg $p$(pressure)-modes, $g$(gravity)-modes, $i$(Coriolis)-modes, $s$(shear)-modes or $w$(wave)-modes. The $f$-mode is the fundamental mode of the $p$-mode sequence and it is the oscillation mode most likely to be excited in violent processes such as neutron star formation by supernova core collapse  (\cite{2018MNRAS.474.5272T, 2019MNRAS.482.3967T}), the pre-merger interaction of neutron stars (see, e.g., \cite{1994MNRAS.270..611L, 1995MNRAS.275..301K, 2011MNRAS.412.1331F, 2012PhRvD..86l1501G, 2017ApJ...837...67C, 2018PhRvD..98j4005C, 2020PhRvD.101h3002S, 
2021MNRAS.506.2985K,2021MNRAS.tmp.2385K}), the early-post-merger oscillations of the final object (\cite{2006PhRvD..73f4027S, 2012PhRvL.108a1101B, 2013PhRvD..88d4026H, 2014PhRvD..89j4021B, 2014PhRvD..90b3002B, 2016CQGra..33r4002L, 2017ApJ...850L..34B, PhysRevD.93.124051, 2015PhRvD..91f4001T, 2018PhRvL.120v1101D, 2019PhRvD.100j4029B}). In the case that the merging neutron stars are of relatively small mass and the post-merger object is a fast spinning neutron star, the unstable $f$-mode oscillations can lead to its spin-down (\cite{2015PhRvD..92j4040D}). The $f$-mode is associated with major density variations and thus can potentially be an emitter of copious amounts of gravitational radiation. The emission of gravitational waves is the primary reason for the mode's rapid damping, at least for newly born neutron stars.

The efforts to associate the patterns of oscillations with the bulk parameters of the stars, \eg their mass, radius or equation of state (henceforth EoS) was initiated in the mid-1990s and continued for almost two decades, advancing the field of gravitational wave asteroseismology (\cite{1996PhRvL..77.4134A, 1998MNRAS.299.1059A, 2001MNRAS.320..307K, 2004PhRvD..70l4015B, 2005PhRvL..95o1101T, 2010ApJ...714.1234L}). To date, very robust empirical relations have been derived for non-rotating neutron stars, connecting observables such as frequency, damping time, or moment of inertia $I$ to the bulk properties; for example, relations of the form $\sigma_0 = \alpha + \beta \sqrt{ M_0/R_0^3 }$ or $M\sigma_0 = F(M_0^3/I)$ (cf. \cite{1996PhRvL..77.4134A, 2010ApJ...714.1234L}) could provide the average density or the moment of inertia of the star if the $f$-mode frequency $\sigma_0$ is known.

In the era of gravitational wave astronomy, the various oscillation patterns (traced already in non-linear numerical simulations, e.g., \cite{2004MNRAS.352.1089S,2006PhRvD..74f4024K,Baiotti08,2009PhRvD..79b4002B,2010PhRvD..81h4055Z,2011MNRAS.418..427S,2015PhRvL.115i1101B}), if observed, can provide a wealth of information about the emitting sources, and their effects can leave their imprints both in the gravitational but also in the electromagnetic spectrum. Moreover, recent studies relate the $f$-mode frequencies to the Love numbers (\cite{2019PhRvC..99d5806W, 2020NatCo..11.2553P, 2021MNRAS.503..533A, 2021PhRvD.104b3005M}) and even to the postmerger short gamma-ray bursts (\cite{2019ApJ...884L..16C}).

%%%%%%%%
\subsection{Rotation}
%%%%%%%%

Nearly all the previously mentioned studies were concerned with non-rotating stars. In nature, neutron stars will always rotate and their rotation rate may reach extreme values. The inclusion of rotation proves difficult since the extreme rotation rates that neutron stars may (and do) reach do not allow to neglect the star's oblateness which removes the spherical symmetry from the system; this in turn makes the mathematical formulation much more involved. As a first approximation, rotation was treated perturbatively, too, which allowed considering even rotating stars as spherically symmetric (\cite{1967ApJ...150.1005H}). In this so-called \emph{slow-rotation approximation}, the perturbation equations gain considerably in complexity and have been written down in a gauge introduced by \cite{1957PhRv..108.1063R} first by \cite{1992PhRvD..46.4289K}. Even though the problem remains one-dimensional, its solution is not straightforward as (among other technicalities) the outgoing-wave boundary condition at infinity is elusive. Notwithstanding,  \cite{1998ApJ...502..708A} successfully applied this formalism and discovered that $r$-modes are prone to the so-called CFS-instability, named after their discoverers \cite{1970PhRvL..24..611C} and \cite{1975ApJ...199L.157F, 1978ApJ...222..281F}, at any rotation rate. There has been continuing effort using the slow-rotation approximation concerning rotational modes (\cite{2001PhRvD..63b4019L, 2003PhRvD..68l4010L}), and also employing different gauges (\cite{2002MNRAS.332..676R, 2008MNRAS.384.1711V, 2008PhRvD..77b4029P}), but with the pressing need for frequencies of \emph{rapidly} rotating neutron stars, the interest slowly faded.

Even though the slow-rotation approximation has proven fruitful in the understanding of neutron star physics, it is no longer applicable when considering neutron stars at arbitrary rotation rates, which is essential for nascent neutron stars or post-merger configurations in the immediate aftermath of a binary merger. Without the spherical symmetry of the problem, one has to account for at least two spatial dimensions which complicates the equations further and amplifies the computational expense; furthermore, it remains elusive how to formulate the outgoing-wave boundary condition at infinity for the spacetime perturbations in two spatial dimensions which essentially removes the possibility to formulate a corresponding eigenvalue problem. This issue can be circumvented by adopting the \emph{Cowling approximation} (\cite{1941MNRAS.101..367C}), in which the spacetime is considered static, also leading to a considerable simplification of the perturbation equations. Ignoring the impact of a dynamic spacetime (which is most severe for the quadrupolar $f$-mode), \cite{1997ApJ...490..779Y, 1999ApJ...515..414Y} computed quadrupolar $f$-mode frequencies of rapidly rotating neutron stars and studied the associated CFS-instability in the late 1990s. \cite{2007PhRvD..75d3007B} revived this approach and investigated the general properties of the spectrum of neutron stars regarding the acoustic and Coriolis-driven modes. As a further step toward a more general relativistic treatment, \cite{2012PhRvD..86j4055Y} revisited the problem in the \emph{conformal flatness approximation}. However, with the mathematical difficulties of extending the eigenvalue formulation to include a dynamic spacetime, the focus shifted to study the oscillation spectra by evolving the perturbation equations in time. 

Despite their complexity, the perturbation equations for rapidly rotating relativistic stars have been written down by \cite{1992MNRAS.254..435P}, even though they were not approached numerically at that time. \cite{2008PhRvD..78f4063G, 2009PhRvD..80f4026G} worked in the Cowling approximation and successfully extracted $f$- and $g$-mode frequencies of arbitrarily uniformly and fast rotating neutron stars and even of differentially rotating ones \cite{2010PhRvD..81h4019K} by adding \emph{artificial viscosity} (also known as \cite{kreiss1973methods} dissipation)  to their evolution equations in order to stabilise their time evolutions. 

During the first decade of the new millennium, substantial advances were made in the time evolution of the unperturbed, non-linear Einstein equations, mostly driven by the aim to simulate compact binary mergers but also applicable to isolated neutron stars. These systems have hardly any symmetries that can be exploited to reduce the complexity of the problem, requiring to carry out the time evolutions on a three-dimensional grid. The upside of which is that essentially no constraints have to be placed on the rotational profile when simulating the dynamics of a neutron star. Such codes have been seen as a promising new approach to the calculation of mode frequencies of rapidly (and differentially) rotating neutron stars and already at the beginning of the decade, the frequencies of axisymmetric modes in the Cowling approximation (\cite{2000MNRAS.313..678F, 2001MNRAS.325.1463F}) and those of (quasi-)radial modes in full general relativity (\cite{2002PhRvD..65h4024F}) had been reported. The non-linear codes kept evolving and were used to generate mode frequencies of $f$-modes in the conformal flatness approximation (\cite{2006MNRAS.368.1609D}) or those of inertial modes in the Cowling approximation (\cite{2008PhRvD..77l4019K}).  \cite{2009PhRvD..79b4002B}. Not much later, the frequencies of non-axisymmetric modes in full general relativity of non-rotating polytropic neutron stars (\cite{2009PhRvD..79b4002B}) and soon those of rapidly rotating polytropic neutron stars (\cite{2010PhRvD..81h4055Z}) were obtained from fully non-linear simulations. Even though successful, this approach to computing the frequencies of non-axisymmetric modes, however, has not been followed closely, which is also due to the computational expense associated with such numerical simulations and the accompanying limited accuracy.

In fact, from the point of view of gravitational wave detectability of oscillation modes, the most relevant scenarios are likely to involve rapidly rotating stars. Unfortunately, the aforementioned empirical relations cannot be trivially extended to rotating stars. Rotation splits the oscillation spectra in a similar fashion as the Zeeman splitting of the spectral lines due to the presence of magnetic fields. In rotating stars, the splitting leads to perturbations propagating in the direction of rotation (so-called \emph{co}-rotating modes) and perturbations traveling in the opposite direction (\emph{counter}-rotating modes). The oscillation frequency as observed by an observer at infinity will either increase or decrease depending on the propagation direction of the waves; for slow rotation there will be a shift of the form $\sigma = \sigma_0 \pm \kappa m \Omega + \mathcal{O}(\Omega^2)$ where $m$ is the angular harmonic index, $\kappa$ a mode and stellar model-dependent constant, and $\Omega$ the angular rotation rate of the star.  If the spin of the star exceeds a critical value, which depends on, \eg the EoS and its mass---\ie when the pattern velocity $\sigma/m$ of the backward moving mode becomes smaller than the star's rotation rate $\Omega/2\pi$---then the star becomes unstable to the emission of gravitational radiation; this is the aforementioned CFS instability. This instability is generic (independent of the degree of rotation) for the $r$-modes (\cite{1998ApJ...502..708A, 1998ApJ...502..714F}) while it can be excited only for relatively high spin values ($\Omega \gtrapprox 0.8 \Omega_K$, with $\Omega_K$ the Kepler velocity) for the quadrupolar $f$-modes. An extensive discussion can be found in \cite{2017LRR....20....7P} and \cite{2018ASSL..457..673G}.

This review is based on the highlights of four recent articles published by the authors, which are \cite{2020PhRvL.125k1106K,2020PhRvD.102f4026K,2021PhRvD.103h3008V,2021PhRvD.104b3005M}.

Throughout this article, we employ units in which $c=G=M_\odot=1$.

\section{Perurbation equations}
\label{sec:formulation}

%%%%%%
\subsection{Background configuration}
%%%%%

We are going to work with the Einstein equations along with the law for the conservation of energy-momentum,
\begin{equation}
    G_{\mu\nu} = 8\pi T_{\mu\nu}
    \quad\text{and}\quad
    \nabla_\mu T^{\mu\nu} = 0,
    \label{eq:Einstein}
\end{equation}
where $G_{\mu\nu}$ is the Einstein tensor and $T_{\mu\nu}$ is the energy-momentum tensor.

We restrict ourselves to the study of the dynamics of small perturbations around an equilibrium configuration which allows us to linearise equations \eqref{eq:Einstein}. We assume an axisymmetric, stationary background configuration for which the metric written in quasi-isotropic coordinates takes the form
\begin{align}
    ds^2
        & = g_{\mu\nu}^{(0)} dx^\mu dx^\nu \\
        & = - e^{2\nu} dt^2 + e^{2\psi} r^2 \sin^2 \theta
            (d\varphi - \omega dt)^2 
      + e^{2\mu} (dr^2 + r^2 d\theta^2).
            \label{eq:metric} \nonumber
\end{align}
Here, $\nu$, $\psi$, $\mu$, and $\omega$ are the four unknown metric potentials, depending only on $r$ and $\theta$.

We model the neutron star to be a perfect fluid without viscosity for which the corresponding energy-momentum tensor  takes the form
\begin{equation}
    T^{\mu\nu} = (\epsilon + p) u^\mu u^\nu + p g^{\mu\nu},
    \label{eq:Energy-Momentum}
\end{equation}
where $\epsilon$ is the energy density, $p$ is the pressure, and $u^\mu$ the 4-velocity of the fluid. The only two non-vanishing components of the 4-velocity are linked via the star's angular rotation rate, $u^\varphi = \Omega u^t$, and by means of the normalisation of the 4-velocity they are given by
\begin{align}
    u^t = \frac{1}{\sqrt{e^{2\nu}
                    - e^{2\psi} \rrsst \left( \Omega-\omega \right)^2}}.
\end{align}
After specifying an EoS, which may be a polytropic or a tabulated one, linking energy density and pressure to each other, we generate uniformly rotating equilibrium configurations using the \texttt{rns}-code (\cite{1995ApJ...444..306S, 1998A&AS..132..431N,rns-v1.1}).

We considered sequences of neutron stars along which we keep either the central energy density constant or the baryon mass fixed (the latter are also known as \emph{evolutionary sequences}) with rotation rates up to their respective mass-shedding limit. Our neutron star models were based on polytropic and realistic EoSs. We considered polytropic models with three different polytropic indices $N = 0.6849, 0.7463$, and $1.0$. Furthermore, we employed piecewise-polytropic approximations, introduced by \cite{2009PhRvD..79l4032R}, for the four tabulated EoSs (APR4, H4, SLy, and WFF1) that we used. Our nonrotating configurations have gravitational masses $M \in [1.17,\,2.19] M_\odot$. Even though current astrophysical constraints play a role in our particular choice of EoSs, it is largely motivated by our desire to provide robust universal relations by covering a wide part of the parameter space.

%%%%%%
\subsection{Perturbation Equations}
%%%%%

As usual in perturbative studies, we decompose the metric as
\begin{align}
    g_{\mu\nu}
        & = g_{\mu\nu}^{(0)} + h_{\mu\nu},
\end{align}
where $g_{\mu\nu}^{(0)}$ is the background metric and $h_{\mu\nu}$ its perturbation. As we will work in the Hilbert gauge, it will be advantageous to work instead with the trace-reversed metric perturbation, defined by
\begin{align}
    \phi_{\mu\nu}
        & := h_{\mu\nu} - \frac{1}{2} g_{\mu\nu}^{(0)} h,
\end{align}
where $h := {h^\mu}_\mu$ is the trace of the metric perturbations.

The metric perturbations are not unique but possess gauge freedom which can be utilised in different ways. Often, the gauge freedom is used to eliminate some of the spacetime perturbations, e.g., by using the well-known gauge by \cite{1957PhRv..108.1063R}, and hence to reduce the number of perturbation equations. In our studies, however, we followed a different approach (which we will reason below) and opt for the Hilbert gauge, which is the gravitational equivalent to the well-known Lorenz gauge in electromagnetism, specified by
\begin{equation}
    f_\mu := \nabla^\nu \phi_{\mu\nu} = 0.
    \label{eq:hilbert_gauge}
\end{equation}
In the Hilbert gauge, the perturbed Einstein tensor takes the form
\begin{align}
    - 2 \delta G_{\mu\nu}
        & = \square \phi_{\mu\nu}
          + 2 R^\alpha{}_\mu{}^\beta{}_\nu \phi_{\alpha\beta}
          + R \phi_{\mu\nu} \nonumber 
        - \left({R^\alpha}_{\mu} \phi_{\nu\alpha}
          + {R^\alpha}_{\nu} \phi_{\mu\alpha}\right)
          - g_{\mu\nu} R^{\alpha\beta} \phi_{\alpha\beta},
\end{align}
where $R^\alpha{}_\mu{}^\beta{}_\nu$, $R^{\alpha\beta}$, and $R$ are the background Riemann tensor, Ricci tensor and scalar curvature, respectively. The advantage of the Hilbert gauge is that the evolution equations for the metric perturbations will take the form of ten coupled wavelike equations (note that in the above expression, the d'Alembert operator, defined with respect to the background metric, is the only differential operator acting on the metric perturbations) while the mixing of temporal and spatial derivatives is avoided. This is in contrast to other common gauge choices (\cite{1998PhRvD..58l4012A, 1971NCimB...3..295B, 2001PhRvD..63f4018R, 2002MNRAS.332..676R}) or the ones without any gauge choice (\cite{1992MNRAS.254..435P}) where the field equations split into subsets of hyperbolic and elliptic equations which have to be either solved simultaneously or by quite cumbersome manipulations to bring them in a hyperbolic form, something that is technically very difficult for the perturbations of rotating stars if not impossible. The fully hyperbolic character of the perturbation equations in the Hilbert gauge makes this gauge particularly convenient for the numerical implementation in a time evolution.

Our choice of gauge, namely the Hilbert gauge, does not eliminate any of the metric perturbations; hence, we need 10 variables to describe the spacetime perturbations. As in previous studies of rotating neutron stars (\cite{PhDVavoul2007, 2010PhRvD..81h4019K}), we use 4 variables for the fluid. In total, we need to evolve 14 variables in time. The evolution equations follow in a very straightforward manner from the perturbed Einstein equations
\begin{align}
    \delta G_{\mu\nu}
        & = 8 \pi \delta T_{\mu\nu},
    \label{eq:pert_Einstein}
\end{align}
and the perturbed law for the conservation of energy-momentum
\begin{align}
    \delta \left( \nabla_\mu T^{\mu\nu} \right) & = 0.
    \label{eq:pert_conslaw}
\end{align}
The evolution equations themselves are quite lengthy and not very enlightening and we refer the reader to \cite{2020PhRvD.102f4026K} for details in the derivation and their implementation.

\section{Results.}

\subsection{Universal Relations for single neutron stars}

As shown in \cite{2020PhRvD.102f4026K}, our code produces results in excellent agreement with previously published values (\cite{2010PhRvD..81h4055Z, 2018PhRvD..98j4005C}) and our convergence tests demonstrate an accuracy of the obtained frequencies of $1 - 2\%$. In this article, we will provide some highlighted results in order to demonstrate the existence of asteroseismological relations of various types and we lay out the way that one can make use of these relations in analyzing gravitational wave signals.

\begin{figure}[htbp]
    \centering
    \includegraphics[width=8.6cm]{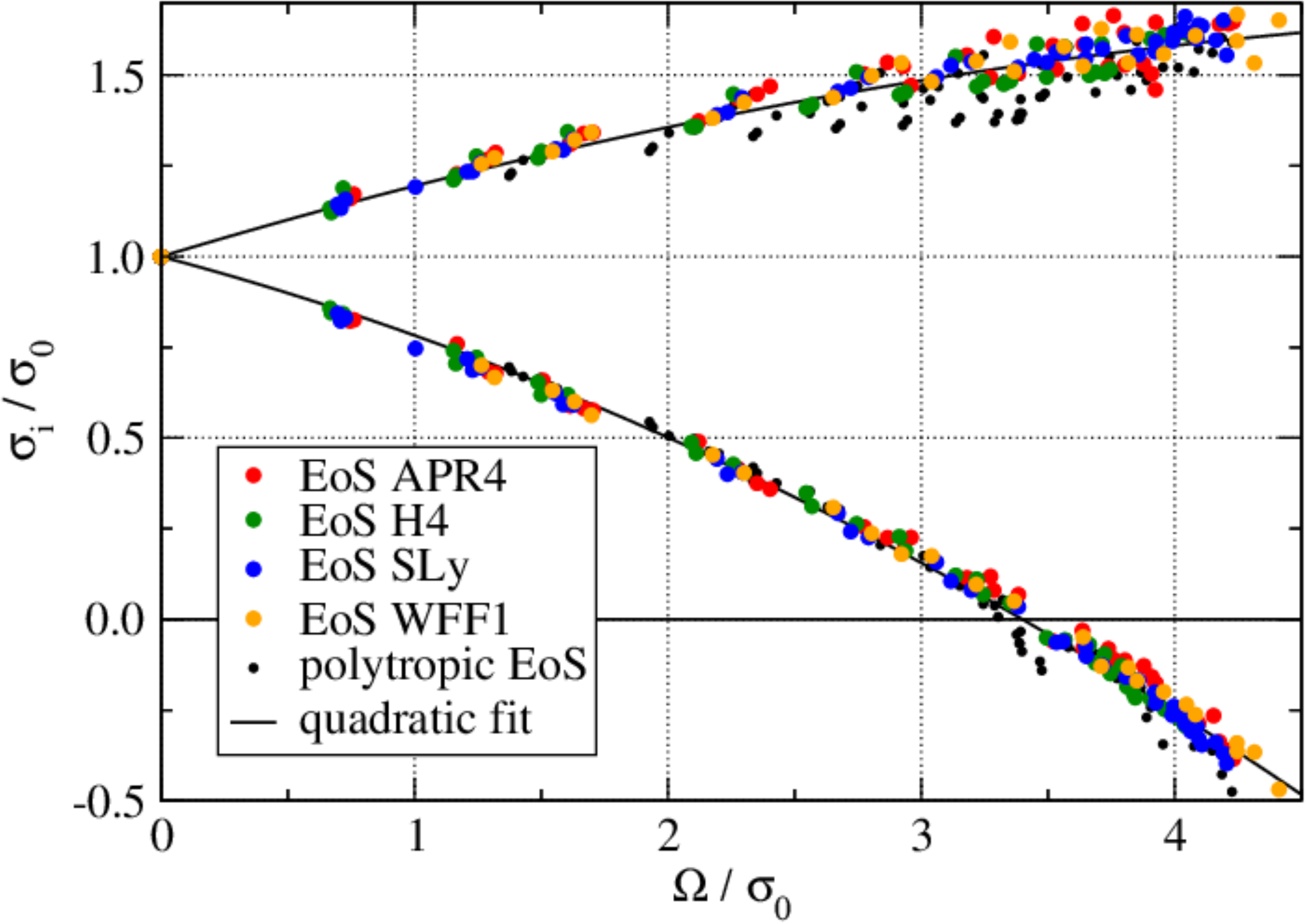}
    \caption{Universal relations for the $l=|m|=2$ $f$-mode frequencies for sequences of constant central energy density as observed in the inertial frame. The graph shows the results from 21 such sequences (three sequences per EoS; three polytropic and four realistic EoSs). The potentially unstable $f$-mode branch displays a strikingly universal behavior; the largest deviations from a quadratic fit occur close to the mass-shedding limit of the sequences, mainly for the polytropic EoSs. Figure taken from \cite{2020PhRvL.125k1106K} with permission by APS.}
    \label{fig:sigma-omega-eps}
\end{figure}

More specifically, we will show different universal relations providing accurate estimates for the $f$-mode frequency given some bulk parameters of the star and vice versa. First, we observe a universal behavior of the $f$-mode frequency $\sigma_\text{i}$ as observed in the inertial frame as a function of the star's angular spinning frequency $\Omega$ along sequences of \emph{fixed central energy density} models when we normalize both frequencies with the $f$-mode frequency $\sigma_0$ of the corresponding non-rotating star. Figure~\ref{fig:sigma-omega-eps} displays this behavior for more than 230 different neutron star models of each the co-rotating (i.e., stable) and counter-rotating (i.e., potentially unstable) branches of the $f$-mode for seven EoSs and various central energy densities (with corresponding central rest mass densities $\rho_c \in [2.2, 7.3] \rho_0$, where $\rho_0 = 2.7 \times 10^{14}\unit{g/cm}^3$ is the nuclear saturation density); we model the universal behavior using the quadratic function
\begin{align}
    \frac{\sigma_\text{i}}{\sigma_0}
        = 1 + a_1 \left( \frac{\Omega}{\sigma_0} \right)
        + a_2 \left( \frac{\Omega}{\sigma_0} \right)^2 .
    \label{eq:omvsom_model}
\end{align}
The results of a least squares fit are $a_1^\text{u} = -0.193$ and $a_2^\text{u} = -0.0294$ for the potentially unstable branch and $a_1^\text{s} = 0.220$ and $a_2^\text{s} = -0.0170$ for the stable branch of the $f$-mode. The quadratic fit accounts well for the increasing oblateness of the star with its rotation; however, close to the Kepler limit, deviations from this simple model become visible. As this deviation is most pronounced for the less realistic polytropic EoSs, we do not take them into account for the quadratic fits. The root mean square of the residuals is $0.024$ for the counter-rotating branch and $0.048$ for the co-rotating branch.

We point out that our model predicts that the unstable branch of the quadrupolar $f$-mode becomes susceptible to the CFS instability once the angular rotation rate of the star exceeds $\Omega \approx \left(3.4 \pm 0.1\right) \sigma_0$ (when considering sequences of constant central energy density); note that the given uncertainty is a bound, not a confidence interval. This finding regarding the critical value complements the well-known threshold of $T/|W| \approx 0.08 \pm 0.01$ in terms of the ratio of rotational to gravitational potential energy (\cite{1998ApJ...492..301S, 1999ApJ...510..854M}), which is confirmed in our simulations and is in contrast to the widely used Newtonian result of $T/|W| \approx 0.14$.

\begin{figure}[htbp]
    \centering
    \includegraphics[width=8.6cm]{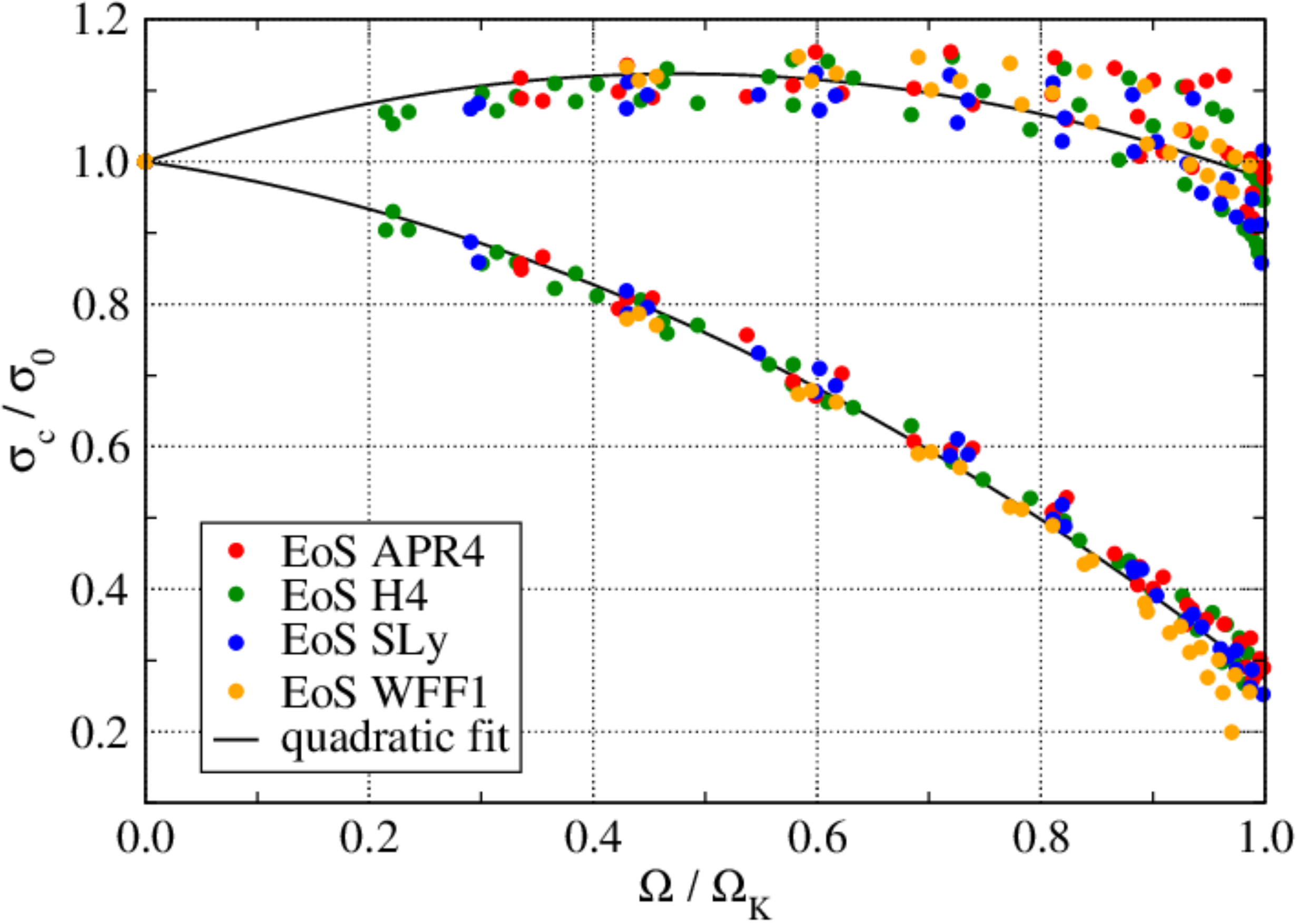}
    \caption{Universal relations for the $l=|m|=2$ $f$-mode frequencies for sequences of constant baryon mass as observed in the comoving frame. The graph shows the results from 12 such sequences (three sequences per EoS). The stable $f$-mode branch displays universal behavior. Figure taken from \cite{2020PhRvL.125k1106K} with permission by APS.}
    \label{fig:sigma-omega-Mo}
\end{figure}

The stable branch of the $f$-mode can be fitted more accurately when switching to the comoving frame and considering sequences of constant baryon mass. The frequency $\sigma_\text{c}$ observed in the comoving frame is related to the frequency observed in the inertial frame via $\sigma_\text{c} = \sigma_\text{i} + m\Omega/2\pi$. We show our results for more than 120 different neutron star models using four realistic EoSs in Figure~\ref{fig:sigma-omega-Mo}. We fit our results to the quadratic function
\begin{align}
    \frac{\sigma_\text{c}}{\sigma_0}
        = 1 + b_1 \left( \frac{\Omega}{\Omega_K} \right)
        + b_2 \left( \frac{\Omega}{\Omega_K} \right)^2 ;
    \label{eq:sigma-omega-Mo_model}
\end{align}
note that we use the Kepler velocity $\Omega_K$ to normalize the star's rotation rate in this formula. The results of a least squares fit are $b_1^\text{u} = 0.517$ and $b_2^\text{u} = -0.542$ for the potentially unstable branch (which in the comoving frame exhibits the higher frequencies) and $b_1^\text{s} = -0.235$ and $b_2^\text{s} = -0.491$ for the stable branch of the $f$-mode. The root mean square of the residuals is $0.024$ for the co-rotating branch and $0.051$ for the counter-rotating branch. \footnote{Similar relations  were presented in \cite{2011PhRvD..83f4031G,2013PhRvD..88d4052D} but in the Cowling approximation.}

In earlier studies for non-rotating models, fitting relations of the form $\sigma_0=\alpha+\beta\sqrt{M_0/R_0^3}$ were derived (\cite{1998MNRAS.299.1059A,2011PhRvD..83f4031G,2013PhRvD..88d4052D}). Here, $\alpha$ and $\beta$ can be estimated for the EoSs that fulfill the constraints at the time of observation while $M_0$ and $R_0$ correspond to the mass and radius of the non-rotating model. Thus, this relation in combination with Eq.~\eqref{eq:omvsom_model} or \eqref{eq:sigma-omega-Mo_model} connects three fundamental parameters of the sequence, \ie mass and radius of the non-rotating member with the spin of the observed model. Obviously, from a single observation of the $f$-mode frequency, one cannot extract these values but can put constraints among the three of them. Any extra observed oscillation frequency, \eg both co- and counter-rotating frequencies or knowledge of some parameters of the star, such as its mass, will place more stringent constraints. 

\begin{figure}[htbp]
    \centering
    \includegraphics[width=8.6cm]{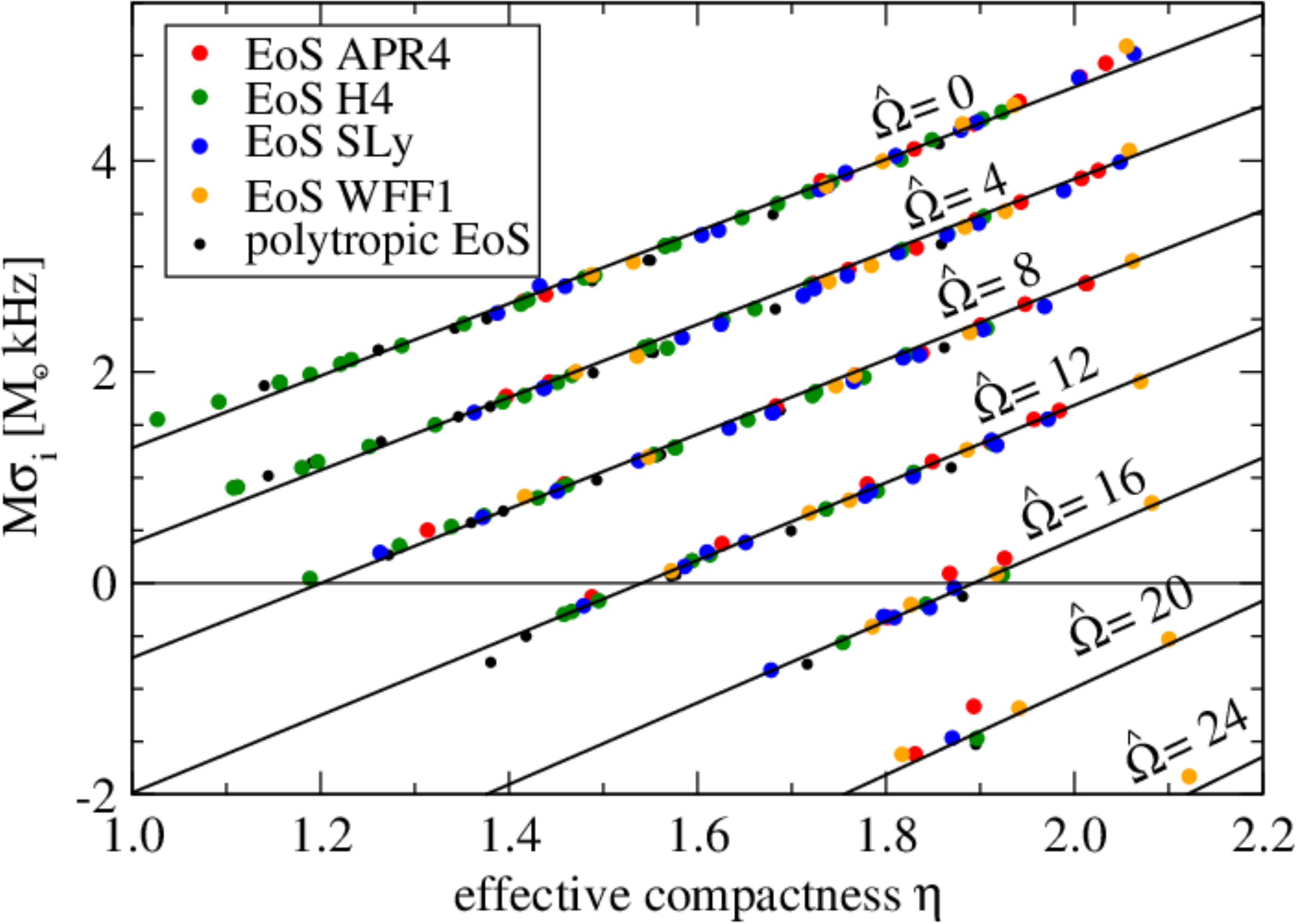}
    \caption{The scaled $f$-mode frequency of the potentially unstable branch in dependence of the effective compactness for different values of $\hat{\Omega} = \bar{M}\Omega/\unit{kHz}$. The straight lines represent the prediction of our fitting formula, cf. Eq.~\eqref{eq:fit-freq-alter}. Figure taken from \cite{2020PhRvL.125k1106K} with permission by APS.}
    \label{fig:eta-Msigma}
\end{figure}

Another fitting relation which can easily be implemented in solving the inverse problem is incorporating the \emph{effective compactness} $\eta := \sqrt{\bar{M}^3/I_{45}}$ (which is closely related to the compactness $M/R$), where $\bar{M}:=M/M_\odot$ and $I_{45}:=I/10^{45}\unit{g}\unit{cm}^2$ are the star's scaled gravitational mass and moment of inertia, inspired by \cite{2010ApJ...714.1234L}. We will be guided by the model employed in the Cowling approximation by \cite{2015PhRvD..92l4004D} which reproduces the $f$-mode frequency of a particular neutron star from its rotation rate, gravitational mass, and effective compactness. We propose the fitting formula
%%%%
\begin{align}
    \hat{\sigma_\text{i}} =
    \left(
        c_1 + c_2 {\hat \Omega} + c_3 {\hat \Omega}^2
    \right) +
    \left(
        d_1 + d_3 {\hat \Omega}^2
    \right) \eta,
    \label{eq:fit-freq-alter}
\end{align}
%%%%
where $\hat{\sigma}_\text{i} := \bar{M}\sigma_\text{i}/\unit{kHz}$ and $\hat{\Omega} := \bar{M}\Omega/\unit{kHz}$; note that we set $d_2 = 0$ as it turns out that this coefficient would be afflicted with a large uncertainty. Using around 100 models based on polytropic as well as around 400 models based on realistic EoSs, the resulting coefficients from a least-squares fit for the counter-rotating branch of the $f$-mode are $(c_1, c_2, c_3)^\text{u}=(-2.14, -0.201, -7.68 \times 10^{-3})$ and $(d_1, d_2, d_3)^\text{u}=(3.42, 0, 1.75 \times 10^{-3})$; for the co-rotating branch, we find the coefficients $(c_1, c_2, c_3)^\text{s}=(-2.14, 0.220, -14.6 \times 10^{-3})$ and $(d_1, d_2, d_3)^\text{s}=(3.42, 0, 6.86 \times 10^{-3})$.
The error in the above reported coefficients is less than 10\,\% and the fitting formula recovers the frequencies with a deviation of less than 20\,\%, with considerably higher accuracy (below 5\,\%) where the $f$-mode frequency is larger than $\approx 500\unit{Hz}$. We show the obtained frequencies along with the predictions from our proposed fitting formula for a few select values of $\hat{\Omega}$, spanning the parameter space up to the Kepler limit, in Fig.~\ref{fig:eta-Msigma}.
 %%%%%%
%%%%%%%
Qualitatively, our coefficients for the counter-rotating branch agree in order of magnitude with those published by \cite{2015PhRvD..92l4004D} in the Cowling approximation; comparing the special case of no rotation, $\hat{\Omega} = 0$, our fitting formula yields roughly $20\,\%$ lower frequencies in our fully general relativistic setup, which is in accordance with expectations.

The lines of constant $\hat{\Omega}$ in Figure~\ref{fig:eta-Msigma} may give the impression that the CFS instability operates more easily in stars with low (effective) compactness, seemingly in contrast to the finding that post-Newtonian effects tend to enhance this instability \cite{1992ApJ...385..630C}. This paradox can be resolved by noting that relativistic effects mainly shift the $f$-mode frequency to lower values while the inclination of the lines of constant $\hat{\Omega}$ is largely unaltered (cf. Figure~3 in \cite{2015PhRvD..92l4004D}). Furthermore, while stars of lower (effective) compactness may reach the neutral point of the $f$-mode indeed at a lower rotation rate, this happens considerably closer to the Kepler limit (if at all) than it would do in more compact stars.

The fitting formula \eqref{eq:fit-freq-alter} has the advantage that it does not rely on specifically defined sequences of neutron stars, along which a particular property is held constant. For example, Eq.~\eqref{eq:sigma-omega-Mo_model} depends on the $f$-mode frequency $\sigma_0$ of the (in a very particular fashion) corresponding non-rotating configuration, which may not even exist in some cases (\eg for supramassive neutron stars supported by rotation); the latter model, cf.~Eq.~\eqref{eq:fit-freq-alter}, is satisfied with bulk properties of the star of which we want to know the oscillation frequency and vice versa. Another benefit of this formulation is that (as demonstrated by \cite{2015PhRvD..92l4004D}) a similar formula can be derived for higher multipoles, \ie $l \ge 3$. Fitting formula \eqref{eq:fit-freq-alter} can be useful in imposing further constraints on the parameters of the postmerger objects since it combines the mass and spin of the resulting object with the $f$-mode frequency and, via $\eta$, the moment of inertia $I$ or the compactness $M/R$. Thus, the latter two can be further constrained by an observation of an $f$-mode signal, as mass and potentially spin can be extracted from the premerger and early postmerger analysis of the signal. The situation becomes more attractive if both co- and counter-rotating modes or other combination of modes are observed since only the mass of the postmerger object will be needed to constrain its parameters by using only the asteroseismological relations (see \cite{2011PhRvD..83f4031G,2015PhRvD..92l4004D,2020PhRvD.101h4039}). This will be an independent yet complementary constraint in the estimation of the radius in addition to those based on the Love numbers (see \cite{2018PhRvD..98h4061D,2018PhRvL.121p1101A,2018PhRvL.121i1102D,2018ApJ...852L..29R}).

%%%%%%%
\begin{figure}[htbp]
    \centering
    \includegraphics[width=8.6cm]{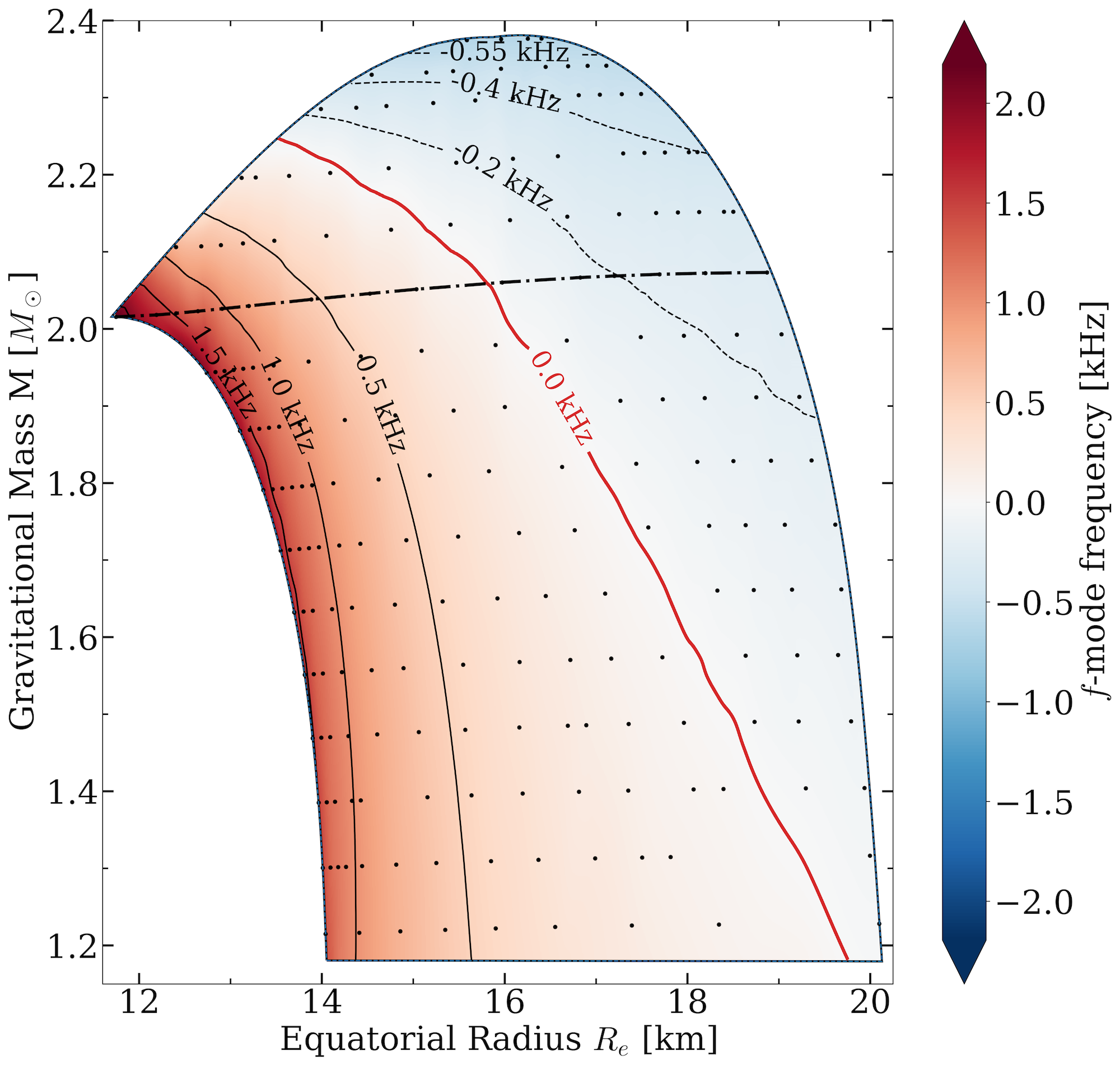}
    \caption{The frequency of the ${}^2f_2$-mode for the entire EoS H4 is displayed color coded and some contour lines are shown. Each black dot indicates a neutron star model for which we have calculated its nonaxisymmetric mode frequencies. All neutron stars located above the (nearly horizontal) dash-dotted line are supramassive. The red (thick) contour line at $0.0\unit{kHz}$ separates the stable models from those that are susceptible to the CFS-instability. Figure taken from \cite{2020PhRvL.125k1106K} with permission by APS.}
    \label{fig:H4fmode}
\end{figure}
%%%%%%%
As a graphical illustration of the behavior of the $f$-mode frequency across the entire parameter space of stable equilibrium models, we present in Figure~\ref{fig:H4fmode} the frequency of the counter-rotating branch as obtained from the time evolutions exemplary for EoS H4; the graph will be qualitatively similar  for other EoSs. We constructed several hundred neutron star models across the entire $M$-$R_e$ plane along so-called \emph{evolutionary sequences}; those are sequences of differently fast rotating neutron stars that share the same baryon mass. The two-dimensional plane of equilibrium models has four distinct boundaries: the static limit at the lower left along which the non-rotating models are located; second, the mass-shedding limit bounds the equatorial radius to the right; next, in the top left of the graph, the limit of stability with respect to quasi-radial perturbations connects the (non-rotating) maximum mass TOV model to (approximately) the heaviest model\footnote{It is well-known that the uniformly rotating equilibrium models with the fastest rotation rate, the highest angular momentum, or the largest gravitational mass may be distinct and it depends on the EoS whether or not those are stable with respect to quasi-radial perturbations (see \cite{1994ApJ...424..823C}).}; equilibrium configurations that are located to the left of this line will collapse to a black hole upon quasi-radial perturbation. Last, in line with current theory and observations (\cite{2018MNRAS.481.3305S,2015ApJ...812..143M}), we limit ourselves to neutron stars with masses $M \gtrsim 1.17 M_\odot$. The dots in Figure \ref{fig:H4fmode} depict the considered equilibrium models. The dash-dotted line slightly above $M = 2.0\,M_\odot$ is the evolutionary sequence with the baryon mass $M_0 = 2.3\,M_\odot$ and it separates the supramassive neutron stars (above that line) which are merely supported by centrifugal force from those (below that line) which may spin down ending up at a stable non-rotating configuration. Other evolutionary sequences can be imagined by combining the dots with lines parallel to the dash-dotted one.

The graph shows that for this particular EoS each neutron star model may become CFS-unstable if it is sufficiently spun up. For heavier neutron stars, this happens considerably below the mass-shedding limit and sufficiently heavy supramassive models (approximately $M \gtrsim 2.25M_\odot$) will inevitably be CFS-unstable (with respect to the quadrupolar $f$-mode); i.e. those stars will be stabilized merely by viscous mechanisms counteracting the CFS-instability.

\subsection{Universal Relations involving long-lived remnants from BNS mergers}

 Inspired by the works  on  universal relations for single neutron stars~(\cite{1998MNRAS.299.1059A,2004PhRvD..70l4015B,2017PhR...681....1Y}),
the last five years have given rise to universal relations for BNS: they relate the pre-merger neutron stars to the early post-merger remnant, and have been developed using numerical
relativity simulations~(\cite{2015PhRvL.115i1101B,2016PhRvD..93l4051R,2020PhRvD.101h4006K}).

These works have primarily focused on relating the tidal deformability of the pre-merger stars, which impact the dynamics of the pre-merger gravitational waves at leading order through the \emph{reduced tidal deformability} or \emph{binary tidal deformability} $\tilde\Lambda$~(\cite{2008PhRvD..77b1502F,2014PhRvL.112j1101F}), to various stellar parameters of the early remnant. 

Recently, \cite{2020PhRvD.101h4039V}  investigated empirical relations for BNS mergers based on the extensive CoRe data set of numerical relativity gravitational wave simulations~(\cite{2018CQGra..35xLT01D}). Covering a wide range of mass ratios, they find an extensive set of universal relations involving the  various peak frequencies of the post-merger gravitational wave signal, involving, e.g., the chirp mass and characteristic radius of a 1.6 $M_\odot$ neutron star.  In particular, they also find universal relations between the binary tidal deformability and the primary $f$-mode frequency of the post-merger signal (as in~\cite{2020PhRvD.101h4006K}), however, this time involving the chirp mass of the BNS. Other established universal relations connect the post-merger peak frequency to the tidal coupling constant (\cite{2015PhRvL.115i1101B, 2019PhRvD.100j4029B}).

The universal relation between the binary tidal deformability of the BNS and the stable, co-rotating $f$-mode frequency $\sigma^s$ of the early, differentially rotating remnant proposed in~\cite{2020PhRvD.101h4006K}, led us in  \cite{2021PhRvD.104b3005M} 
to derive a similar relation for a potentially long-lived, uniformly rotating remnant: the relation takes the form
\begin{equation}
    \log_{10} \hat\sigma^s = a(q) \cdot \tilde \Lambda^{\frac{1}{5}} + b(q),
\label{eq:indirect1}
\end{equation}
where $\hat \sigma^s = \frac{M}{M_\odot}\frac{\sigma^s}{\text{kHz}}$ is the normalized co-rotating $f$-mode frequency, and $q = \frac{M_1}{M_2} \leq 1$ the gravitational mass ratio of the pre-merger stars. For rapidly rotating, long-lived remnants (with rotation frequency $\bar\Omega \geq 800\,\text{Hz}$), this relation achieves an average relative error of $1.3\,\%$.

We also derive a relation for the potentially unstable, counter-rotating $f$-mode frequency of the long-lived remnant, presenting the possibility of predicting the onset of the earlier mentioned CFS-instability.

Combining these results with the universal relation, Eq.~\eqref{eq:sigma-omega-Mo_model}, for fast rotating neutron stars   between the stable, co-rotating $f$-mode frequency and the effective compactness $\eta = \sqrt{\bar M^3/I_{45}}$ (as defined in the previous section),
we also derived a combined relation of the form
\begin{equation}
    \eta = \frac{10^{a(q) \cdot \tilde \Lambda^{\frac{1}{5}} + b(q)} - \left(c_1 + c_2 \hat \Omega + c_3 \hat \Omega^2\right)}{d_1 + d_3 \hat \Omega}
\label{eq:indirect2}
\end{equation}
that relates the pre-merger binary tidal deformability of the BNS with the effective compactness of the long-lived remnant. For rapidly rotating remnants, this relation achieves an average relative error of $2.4\,\%$.

\begin{figure}[htbp]
    \centering
    \includegraphics[width=8.6cm]{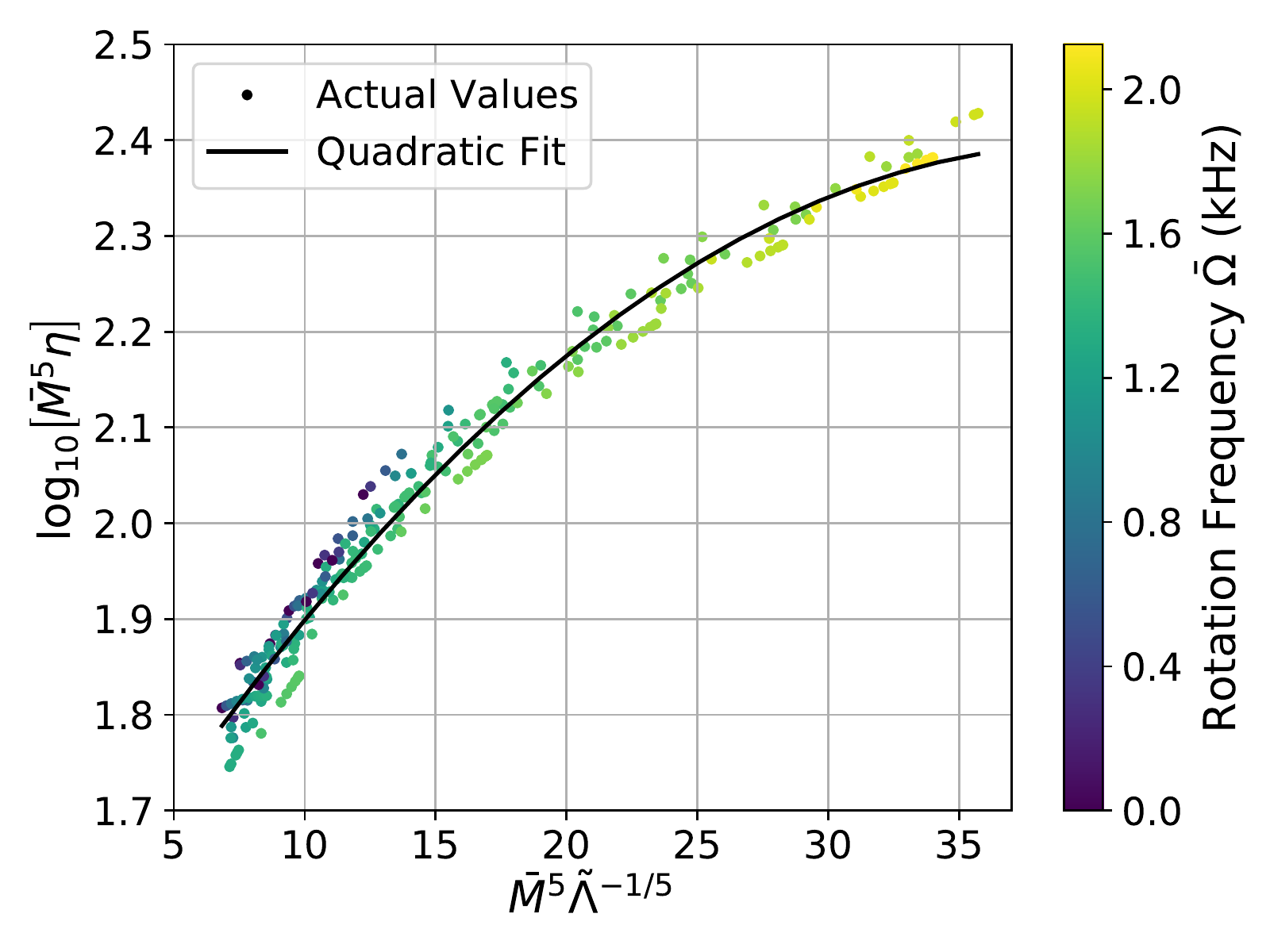}
    \caption{The quadratic fit between $\bar{M}^5 \tilde{\Lambda}^{-1/5}$ and $\bar{M}^5C$ for $q = 1$. The fit corresponds to Eq.~\eqref{eq:direct}. Figure taken from \cite{2021PhRvD.104b3005M} with permission by APS.}
    \label{fig:prav-quadratic}
\end{figure}

Finally, by directly relating these quantities without going via the $f$-mode, we obtain a universal relation of the form
\begin{equation}
    \log\left[\bar M^5 \eta\right] = a(q) \left(\bar M^{5}\tilde\Lambda^{-\frac{1}{5}}\right)^2 + b(q) \bar M^{5}\tilde\Lambda^{-\frac{1}{5}}+ c(q).
    \label{eq:direct}
\end{equation}
This relation achieves improved accuracy, reaching an average relative error of $\sim 1.5\,\%$ for remnants with any rotation frequency. We show the quadratic fit in Fig.~\ref{fig:prav-quadratic} for the symmetric case of $q = 1$.

\begin{figure}[htbp]
    \centering
    \includegraphics[width=8.6cm]{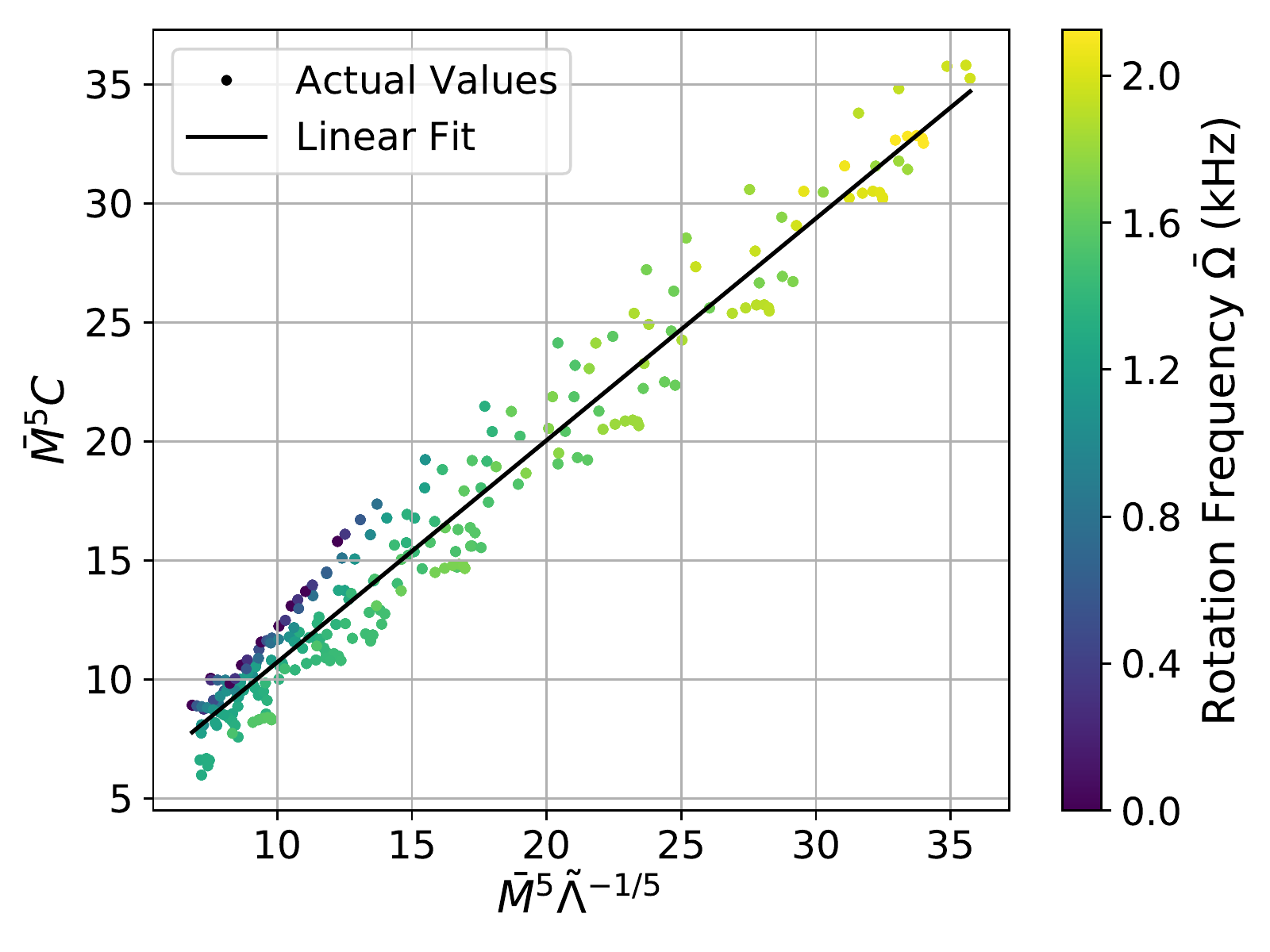}
    \caption{The linear fit between $\bar{M}^5 \tilde{\Lambda}^{-1/5}$ and $\bar{M}^5C$ for $q = 1$, considering only soft EoSs. The fit corresponds to Eq.~\eqref{eq:direct2}. Figure taken from \cite{2021PhRvD.104b3005M} with permission by APS.}
    \label{fig:prav-compactness}
\end{figure}

We finally also consider a direct relation between the binary tidal deformability and the compactness $C = M/R$ of the long-lived remnant. 
Such a relation would allow the direct estimation of the remnant's radius $R$ using independent estimates of its gravitational mass. We propose a relation of the form
\begin{equation}
    \bar M^5 C = a(q) \bar M^{5} \tilde \Lambda^{-\frac{1}{5}} + b(q)
    \label{eq:direct2}
\end{equation}
which, however, only achieves an accuracy an order of magnitude worse than for the effective compactness relation, reaching an average relative error of $\sim 8.8\,\%$. A graphic representation of this fit along with the data is shown in Figure~\eqref{fig:prav-compactness}.

The results presented in \cite{2021PhRvD.104b3005M} represent a first step towards finding universal relations between the pre-merger neutron stars and the potential long-lived remnant 
of a BNS merger using perturbative calculations. Our approach can be freely extended to, e.g., hot EoSs, phase transitions, as well as differential rotation for the remnant to, to cover, e.g., earlier parts of the post-merger phase. 

The various functional expressions for $a(q)$, $b(q)$ and $c(q)$ used in equations
(\ref{eq:indirect1}),
(\ref{eq:indirect2}),
(\ref{eq:direct}),
(\ref{eq:direct2}) as well as the coefficients $c_1$, $c_2$, $c_3$, $d_1$ and $d_2$ are given in \cite{2021PhRvD.104b3005M}.

\section{Approaching the inverse problem}

Having laid out the procedure to extract fluid oscillation frequencies from time evolution in a general relativistic framework and having developed various universal relations between bulk properties of neutron stars and their $f$-mode frequencies, we now turn to one of the possible applications. The field of gravitational wave asteroseismology attempts to invert the above outlined procedure and aims to constrain bulk properties of neutron stars such as mass and radius given one or more oscillations frequencies. The above derived universal relations are a valuable tool for this as they do not depend on the underlying and hitherto poorly understood nuclear equation of state. However, even without universal relations at hand, the inverse problem can be approached.

We will summarise an approach of the inverse problem using Bayesian methods as laid out by \cite{2021PhRvD.103h3008V}. Here, the fundamental idea is tested on the $f$-mode frequency, but it may easily be extended to other fluid modes or combined with their damping times. The use of the Bayesian framework is advantageous (compared to, e.g., an analytical inversion of the non-linear universal relation) as it directly allows to incorporate error bars of the measurement of the $f$-mode into the calculation. Besides the Bayesian approach and universal relations, there are in principle also semi-classical techniques, e.g., WKB theory, which can be used to address the inverse spectrum problem in simplified cases; see \cite{Volkel:2019gpq}¸for using axial $w$-modes of spherically symmetric neutron stars and \cite{Volkel:2017kfj} using a similar approach for ultra compact stars.

\begin{figure}[htbp]
    \centering
    \includegraphics[width=8.6cm]{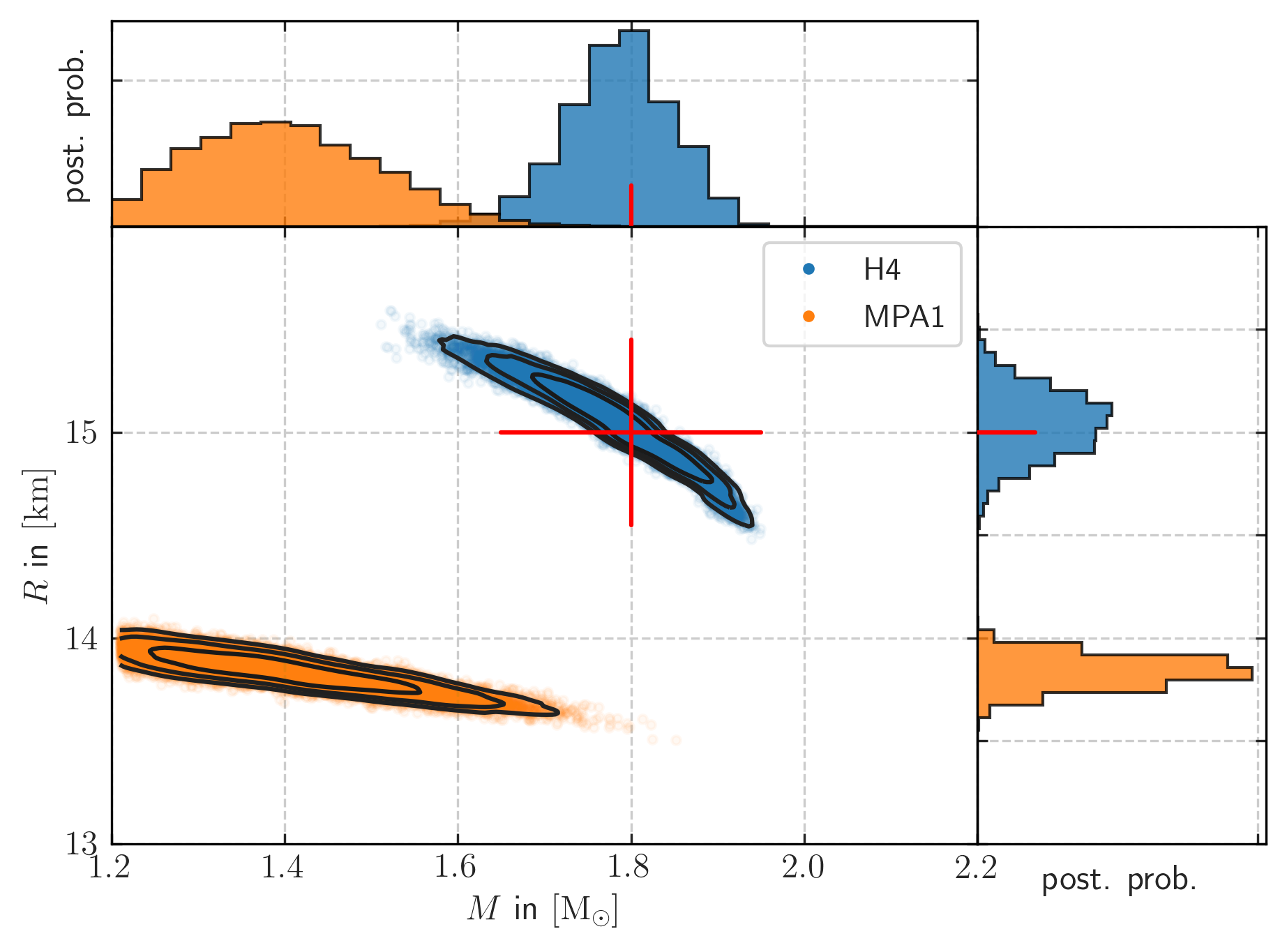}
    \caption{Here we compare the EoS method assuming the H4 EoS (blue) and the MPA1 EoS (orange). The diagonal panels show the sampled posterior distribution of $M$ and $R$, while the off diagonal panel combines a scatter plot with logarithmic contour lines. The red cross and red line indicate the true H4 parameters that belong to the assumed $3\,\%$ $f$-mode data. For both cases we assume that the mass is known by $10\,\%$. Figure taken from \cite{2021PhRvD.103h3008V} with permission by APS.}
    \label{fig:seb-eos_methods}
\end{figure}

In a first step, we pick an EoS and a particular (rotating) neutron star model for which we then calculate two $f$-mode frequencies. We assume those to have a relative error of $3\,\%$ (Gaussian), while our prior knowledge on $M$ and $R$ is uninformative. Two data points should in principle suffice to uniquely pinpoint one neutron star model (assuming a cold EoS). We show the result of the Bayesian analysis in Figure~\ref{fig:seb-eos_methods}. The initially chosen model is depicted by the red cross at $M = 1.8\,M_\odot$ and $R = 15\,\text{km}$. The blue shaded area in the big panel shows the correlations of mass and radius during sampling, whose posterior distributions are shown in the side panels. While the radius of the star can be reconstructed with fairly small error bars, the reconstructed mass is considerably less constrained. Nonetheless, the peak of the probability distribution nicely resembles the initially chosen model. If the $f$-mode frequencies were to be known more accurately, the corresponding error bars of reconstructed mass and radius would be smaller, too.

This analysis required the assumption of a particular EoS and so far we used the same EoS to reconstruct the model that we also used to generate the underlying neutron star model. If we use a different EoS to perform the Bayesian analysis, we will obviously reconstruct a different model; this is shown in orange in Figure~\ref{fig:seb-eos_methods}. The reconstructed star in this case is considerably smaller and lighter. Note that the apparent cut of the MPA1 EoS posteriors (in particular for the mass) is not an indication that this EoS has not been used for the injection, since a similar behavior can also be found for the H4 EoS, if the injected parameters are closer to the edge of the H4 EoS neutron star parameter space. One needs further information, such as the precise mass, the tidal deformability, or perhaps a damping time, in order to rule out this EoS.

Instead of making an educated guess for the ``correct'' EoS and reconstructing mass and radius as just described, we may also use the universal relation proposed in Eq. \eqref{eq:fit-freq-alter}. We will not need to assume any EoS but the universal relation will provide us---given two $f$-mode frequencies and a good estimate of the star's mass---with an estimate for the stars's rotation rate $\Omega$ and its effective compactness $\eta$.

\begin{figure}[htbp]
    \centering
    \includegraphics[width=8.6cm]{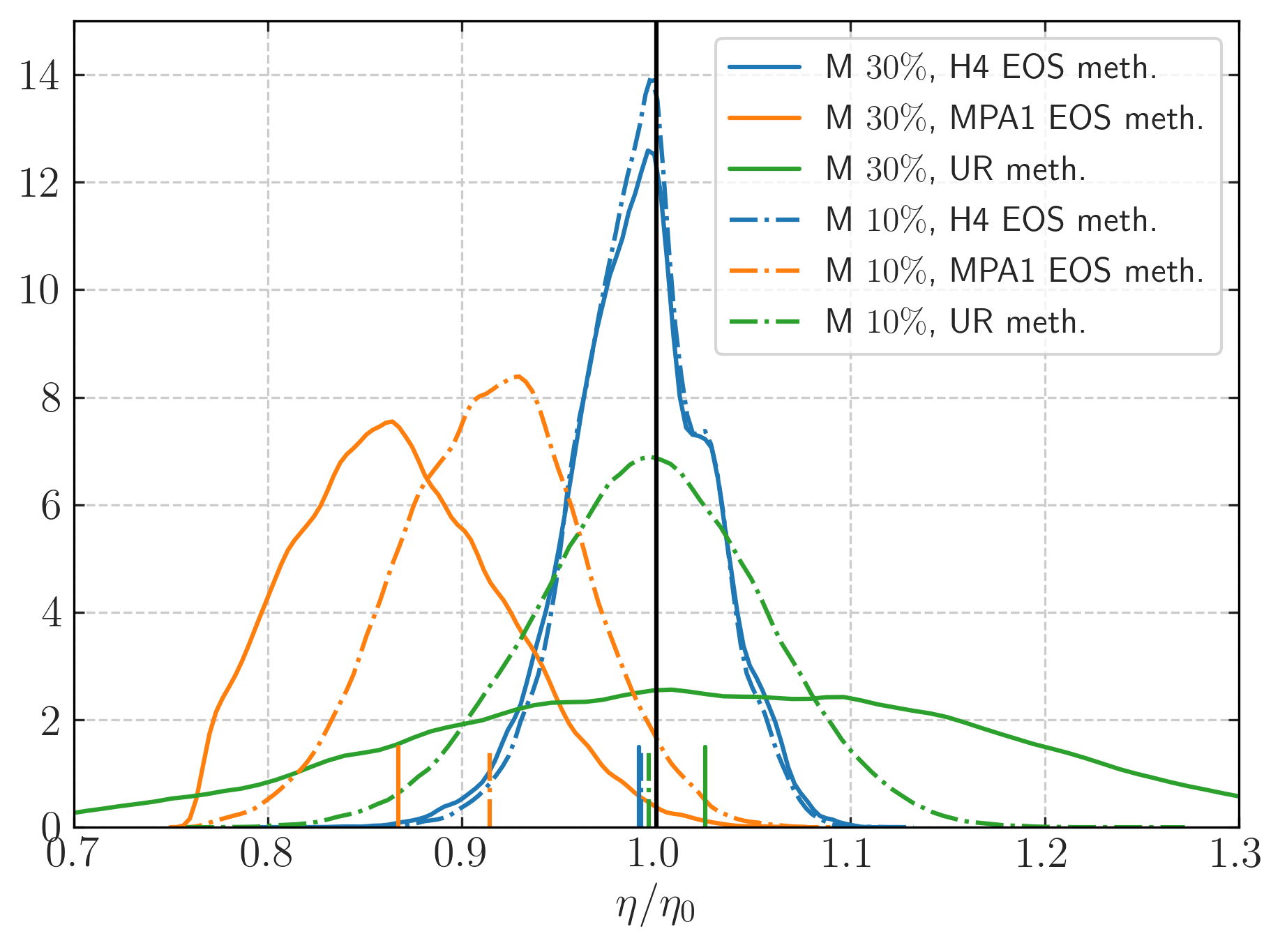}
    \includegraphics[width=8.6cm]{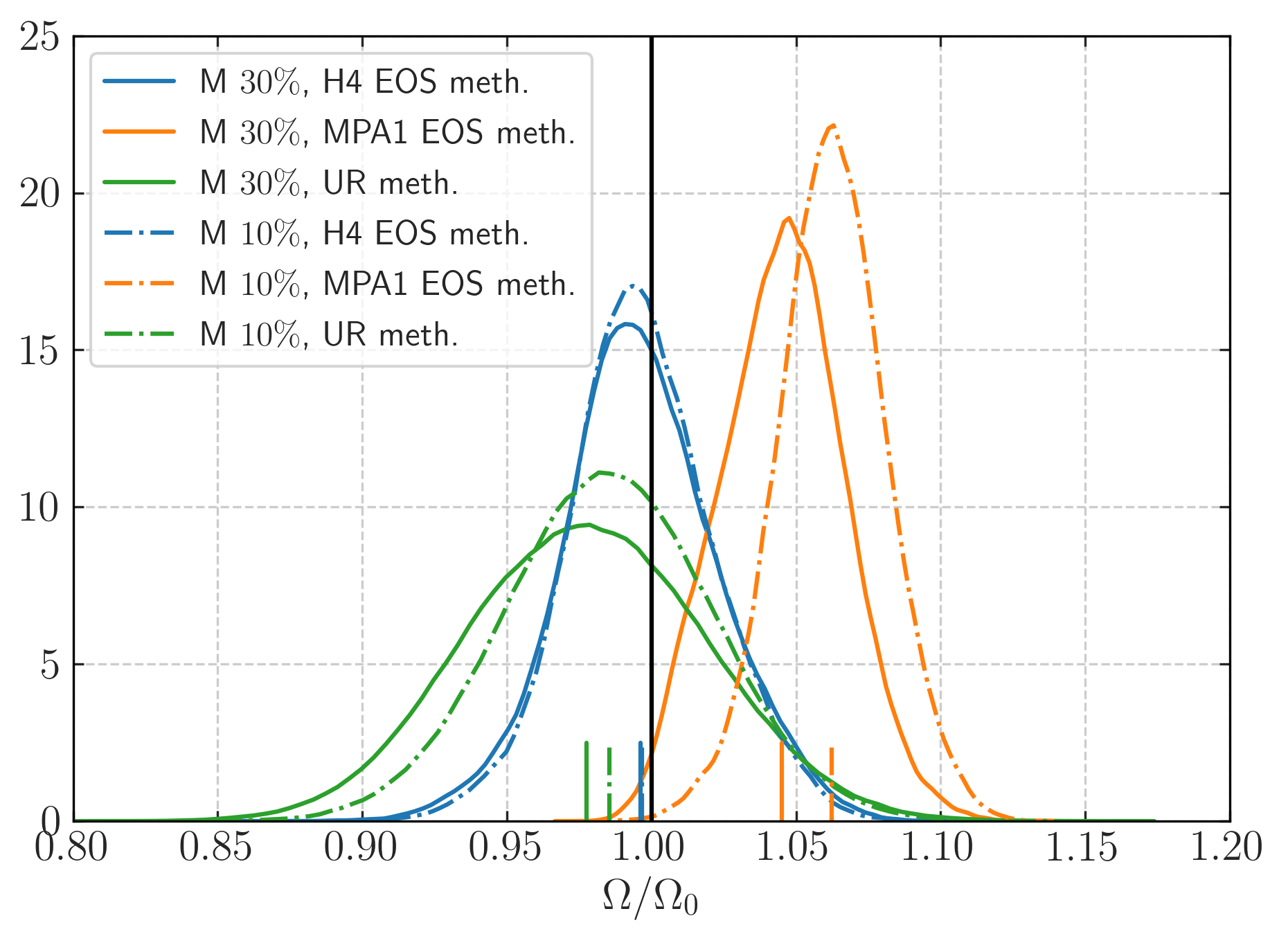}
    \caption{Here we show the posterior distributions of the rotation rate $\Omega$ (top panel) and effective compactness $\eta$ (bottom panel) normalized to the injected H4 values ($\Omega_0$,$\eta_0$). Posteriors are obtained by using the EoS method with H4 EoS (blue), the MPA1 EoS (orange), as well as the universal relation method (green). Solid lines correspond to 30\,\% relative error on the prior mass M and dashed lines to 10\,\%. We indicate each mean of the posteriors as vertical lines. The $f$-mode relative error is assumed to be 3\,\%. Figure taken from \cite{2021PhRvD.103h3008V} with permission by APS.}
    \label{fig:seb-eos_vs_ur_omega}
\end{figure}

We again use the Bayesian method to invert the universal relation and we apply it to the same model as in the previous analysis; the H4 model with $M = 1.8\,M_\odot$ and $R = 15\,\text{km}$. In Figure~\ref{fig:seb-eos_vs_ur_omega} we show the results of the analysis. On the $x$-axis, we show the star's effective compactness and rotation rate, both normalised to the actual value corresponding to the equilibrium configuration. The two graphs then show the posterior distributions. We ran the same analysis twice, once assuming a $30\,\%$ relative error on the prior mass (solid lines) and once assuming a $10\,\%$ relative error (dashed lines). As expected, with a less informed prior on the mass, the posterior distribution for the desired quantities have a considerably larger variance.

The blue and orange curves correspond to the previously describe method where we assume a particular EoS. We have described this above and here we show the posteriors for $\eta$ and $\Omega$. The green curves show the posterior distribution employing the universal relation.

It is evident that the H4 EoS method (blue lines) and universal relation (UR) method (green lines) yield very similar results for the rotation rate $\Omega$,
while assuming the MPA1 EoS (orange lines) indicates a value that is larger than the correct one. Note that both observations hold independent
of the specific prior knowledge of $M$ assumed here ($30\,\%$ and $10\,\%$).

The situation for the effective compactness $\eta$ is qualitatively different. First, the prior knowledge of $M$ plays a big role for the UR method, but is less important for the EoS methods. For those we find that the correctly assumed H4 EoS is almost independent of uncertainties in $M$, while the posterior distribution obtained by the MPA1 EoS is shifted. Note that the rather different scaling behavior of the UR method is in agreement with the findings described above.

Finally, while the posteriors of $\Omega$ are very smooth, one observes small
``bumps'' for the H4 EoS, e.g., at $\eta/\eta_0 \approx 1.02$. We have verified that this likely is an artifact from the finite resolution and particular range of the used H4 $f$-mode data that is available to us. This directly sets the scale of how precise our currently implemented EoS data can be used to resolve the underlying parameters, which is of order percent level.

\section{Summary and Outlook}

We report the first extraction of frequencies of the $l=|m|=2$ $f$-mode of general relativistic, rapidly rotating neutron stars \emph{without} the commonly used \emph{slow-rotation} or \emph{Cowling approximation} to an extent that allows us to generalize the findings into universal relations. This concludes a long-standing open problem, building upon the effort from numerous studies throughout the past five decades. 

For this, we have derived a set of time evolution equations governing the perturbations of the fluid of a rapidly rotating neutron star as well as its surrounding spacetime, derived in a perturbative framework in full general relativity. We have opted for the Hilbert gauge in order to arrive at a set of fully hyperbolic equations for the spacetime perturbations whose implementation does not pose major obstacles.

Convergence tests to study the accuracy of our code reveal that in a typical simulation, the obtained frequency usually deviates less than $1\,\% - 2\,\%$ from the limiting value or has an accuracy of about $10\unit{Hz}$. As we model axisymmetric configurations only, we can reduce the problem to two spatial dimensions which drastically lowers the computational expense of our numerical time evolutions in comparison to non-linear codes that perform three-dimensional simulations. The evolution of the perturbations of a neutron star for $15\unit{ms}$ on a grid with $3120 \times 50$ points, which facilitates a decent frequency resolution, requires only a few dozen of CPU hours; this enables us to study broad ranges of parameters and various EoSs. We expect to further reduce the computational cost even further by deriving a simplified set of perturbation equations.

As a result, we provide different universal relations for the frequencies of the $l=|m|=2$ $f$-modes of uniformly rotating neutron stars at zero temperature which are independent of the EoS; the proposed formulae are calibrated to several hundred neutron star models that are constructed using both polytropic and realistic EoSs and are scattered across the entire parameter space of equilibrium solutions. Furthermore, it is possible to link the pre-merger binary tidal deformability to the effective compactness of the late post-merger remnant in a universal manner, too. Such universal relations will be an essential piece in the asteroseismological toolkit once the third-generation GW observatories will be able to pick up the ring-down and fluid ringing signal following the merger of a binary neutron star system; they allow to solve the inverse problem, leading to significantly tighter constraints for mass and radius of the postmerger object. For this task, it is elementary to have a smorgasbord of universal relations at hand, which allows to make a practical choice, depending on which observables are available or in which of the star's properties one is interested. We will extend the present list of such universal relations in future articles, utilizing different combinations of bulk properties of the star; while we may obviously be (and already have been) inspired by previously published fitting formulae that were derived using different approximative frameworks, we need to be open-minded about models involving novel combinations of observables.

We also report the discovery of an accurate estimate for the onset of the CFS-instability when the $f$-mode frequency of the non-rotating member of the family is known and verified the corresponding critical value of $T/|W|$.

A natural extension of the present study will be a more comprehensive investigation of the spectrum of neutron stars (i.e. higher multipole $f$-modes, low $p$-modes and $g$-modes as well as $w$-modes) which may be excited in different astrophysical processes. Furthermore, we are going to extend our code to account for differentially rotating neutron stars and hot EoSs that are particularly relevant for nascent neutron stars or postmerger configurations in the immediate aftermath of a binary merger, both of which will have a considerable impact on the vibration frequencies or the onset of the CFS-instability (and via two further scaling parameters also on the universal relations) during a very short but dynamic interval of their lives.

Furthermore, a precise knowledge of the spectrum of compact objects is invaluable as also isolated neutron stars as well as those in inspiraling binary systems may possess high spin rates and various oscillation modes, which may be excited, e.g., via glitches or tidal coupling, may impact their electromagnetic emission or the dynamic of the whole system.

\section{Acknowledgements}
C.K. acknowledges financial support through DFG research Grant No. 413873357. S.V. acknowledges financial support provided under the European Union's H2020 ERC Consolidator Grant ``GRavity from Astrophysical to Microscopic Scales'' grant agreement no. GRAMS-815673. A part of the computations were performed on Trillian, a Cray XE6m-200 supercomputer at UNH supported by the NSF MRI program under Grant No. PHY-1229408.

\end{document}